\documentstyle[psfig]{mn}

\newif\ifAMStwofonts

\newcommand{\lapp}{\mbox{\raisebox{-0.3em}{$\stackrel{\textstyle <}{\sim}$}}}
\newcommand{\gapp}{\mbox{\raisebox{-0.3em}{$\stackrel{\textstyle >}{\sim}$}}}

\title{Radio bubbles in the composite AGN-starburst galaxy NGC~6764}
\author[Ananda Hota and D.J. Saikia]
       {Ananda Hota$^1$,$^2$\thanks{hota@ncra.tifr.res.in} and D.J. Saikia$^2$\thanks{djs@ncra.tifr.res.in} \\
$^1$ Joint Astronomy Programme, Indian Institute of Science, Bangalore 560 012, India \\
$^2$ National Centre for Radio Astrophysics, TIFR, Pune University Campus, Post Bag 3, 
Pune 411 007, India}
\date{Accepted.    Received }

\pagerange{\pageref{firstpage}--\pageref{lastpage}}
\pubyear{  }

\begin{document}

\maketitle

\label{firstpage}

\begin{abstract}
We present multi-frequency radio continuum as well as \hbox{H\,{\sc i}}
observations of the composite galaxy NGC~6764, which has a young, circumnuclear
starburst and also harbours an active galactic nucleus (AGN).
These observations have been made at a number of frequencies ranging
from $\sim$600 MHz to 15 GHz using both the Giant Metrewave Radio Telescope 
(GMRT) and the Very Large Array (VLA). They reveal the structure
of the bipolar bubbles of non-thermal, radio emission which are along
the minor axis of the galaxy and extend up to $\sim$1.1 and 1.5 kpc on the
northern and southern sides respectively. Features in the radio bubbles appear to 
overlap with filaments of H$\alpha$ emission. The high-resolution observations
reveal a compact source, likely to be associated with the nucleus of the
galaxy, and a possible radio jet towards the south-west. We have compiled
a representative sample of galaxies with bubbles of non-thermal radio
emission and find that these are found in galaxies with an AGN. 

The \hbox{H\,{\sc i}} observations with the GMRT show two peaks of
emission on both ends of the stellar-bar and depletion of \hbox{H\,{\sc i}}
in the central region of the galaxy.  We also detect \hbox{H\,{\sc i}} in absorption 
against the central radio peak at the systemic velocity of the galaxy. The
\hbox{H\,{\sc i}}-absorption spectrum also suggests a possible weak absorption 
feature blue-shifted by $\sim$120 km s$^{-1}$, which requires confirmation.
A similar feature has also been reported from observations of CO in emission, suggesting
that the circumnuclear starburst and nuclear activity affect the kinematics of the atomic and   
molecular gas components, in addition to the ionised gas seen in H$\alpha$ 
and [N{\sc ii}].
\end{abstract}


\begin{table*}
 \centering
 \begin{minipage}{140mm}
  \caption{Basic data on NGC~6764.$^a$}
  \begin{tabular}{@{}cccccccc@{}}
\hline
 RA (J2000)$^b$
& DEC (J2000)$^b$ & Type$^c$ &
 a $\times$ b$^d$ & V$_{sys}$$^e$ & i$^f$ & log(L$_{FIR}$)$^g$ & D$^h$ \\
 (h m s) & ($^\circ$ $^{\prime}$ $^{\prime\prime}$) &
& ($^\prime$ $\times$
$^\prime$) & (km s$^{-1}$) & ($^\circ$) & (L$_\odot$) & (Mpc) \\
\hline
 19 08 16.428  & 50 55 59.47   & SB(s)bc &
2.3 $\times$ 1.3 & 2416 $\pm$4 & 44.5 & 10.31 & 34.0 \\
\hline
\end{tabular}\hfill\break
$a$ Taken from the NASA Extragalactic Database (NED), unless stated otherwise.
\hfill\break
$b$ The position of radio peak from our high-resolution, VLA A-array, 8460-MHz image. \hfill\break
$c$ Morphological type.\hfill\break
$d$ Optical major and minor axes.\hfill\break
$e$ Heliocentric systemic velocity. \hfill\break
$f$ Inclination angle from H{\sc i} study by Wilcots, Turnbull \& Brinks (2001). \hfill\break
$g$ Log of the far infra-red luminosity (Condon et al. 1996) revised for a distance of 34.0 Mpc. \hfill\break
$h$ Distance estimated using the galaxy recessional velocity and  H$_0$=71 km s$^{-1}$ Mpc$^{-1}$ 
    (Spergel et al. 2003) using the web-based cosmology calculator of Ned Wright. 
    For this distance 1$^{\prime\prime}$=163 pc.
\end{minipage}
\end{table*}

\section{Introduction}
Galactic-scale outflows or galactic winds play an important role in heating and 
supplying kinetic energy to the intergalactic medium (IGM) and enriching it with 
metals (e.g. Heckman, Armus \& Miley 1990; Veilleux, Cecil \& Bland-Hawthorn 2005 
for a review). 
These outflows can be driven either by a circumnuclear starburst or an active
galactic nucleus (AGN), although it is often likely to be due to a combination of
both these processes. The conditions for an AGN, such as a deep potential well
and a supply of gas, could also lead to a circumnuclear starburst and trigger a
starburst-driven galactic wind. Observationally, about half of nearby Seyfert 2 
galaxies are found to host a nuclear starburst, with the proportion being larger for
infra-red selected objects (e.g. Cid Fernandes et al. 2001; Veilleux 2001 and 
references therein). Also in some cases, such as in NGC3079, there is strong
evidence that starburst-driven winds co-exist with the AGN-driven outflows
(Cecil et al. 2001; Irwin \& Saikia 2003).  
Although it is usually not straightforward to disentangle the 
contributions to the galactic-scale outflows from the AGN and the starburst, the galactic 
winds from starbursts tend to be oriented along the minor axes of their parent 
galaxies, while the outflows from AGN show no preferred orientation (e.g Ulvestad
\& Wilson 1984; Kinney et al. 2000; Gallimore et al. 2006 and references therein).  

In this paper, we present radio continuum and H{\sc i} observations of the S-shaped
spiral galaxy, NGC6764, which hosts an AGN and also exhibits evidence of a recent, intense
starburst. It has been classified as a Seyfert 2 galaxy (Rubin, Thonnard \& Ford 1975)
and also as a low-ionization nuclear emission line region or LINER (Osterbrock \& Cohen 1982;
Schinnerer, Eckart \& Boller 2000;  Eckart et al. 1991, 1996). X-ray observations
exhibit evidence of variability by a factor of $\sim$2 over a time scale of 7 days,
consistent with the existence of an AGN (Schinnerer et al. 2000). The basic properties of the 
galaxy are summarised in Table 1. 
It is strongly barred and has a compact nuclear region, bright in both optical
emission lines (Eckart et al. 1996) and radio continuum (Baum et al. 1993). There is 
a very prominent 466 nm Wolf-Rayet (W-R) emission feature from the nucleus (Osterbrock
\& Cohen 1982; Schinnerer et al. 2000), suggestive of recent star formation 
(Armus, Heckman \& Miley 1988; Conti 1991). There is a dense concentration of molecular 
gas in the central region 
(Eckart et al. 1991, 1996; Kohno et al. 2001) with T$_{\rm kin}>$ 20K in the
circumnuclear region and $\lapp$10K towards the spiral arms (Eckart et al. 1991). Interferometric
observations show molecular cloud complexes with an asymmetric velocity field
which is blue-shifted relative to the systemic velocity. The distribution of the 
2.12 $\mu$m H$_2$ line flux is orthogonal to the stellar bar and also indicates 
possible outflow from the circumnuclear region (Eckart et al. 1996). The radio emission
is extended roughly orthogonal to the major axis, forming bubbles of emission on
opposite sides. There is also an extended X-ray component similar in extent to that
of the radio emission (Schinnerer et al. 2000). 
  
The new radio continuum observations presented here were made with the Very Large 
Array (VLA) and the Giant Metrewave Radio Telescope (GMRT) at a number of frequencies
ranging from $\sim$600 MHz to 15 GHz. These observations reveal the structure of the
bubbles of non-thermal emission extending up to $\sim$7 and 9 arcsec on the
northern and southern sides respectively, as well as
that of the circumnuclear region, in far greater detail. We present spectral index images 
of these features, compare NGC6764 with other sources with bubbles of non-thermal
plasma and discuss the possible origin of these features. The H{\sc i} observations
were made with the GMRT to examine the distribution of H{\sc i} clouds with higher
resolution and also possible kinematic effects of the circumnuclear starburst on
the H{\sc i} clouds via H{\sc i}-absorption spectra. We describe the observations
in Section 2, and present the observational results from our radio continuum and H{\sc i} 
observations in Sections 3 and 4 respectively. These are discussed
in Section 5 while the conclusions are summarised in Section 6.             

\section{Observations and Data analysis}
The observing log for both the GMRT and VLA observations is presented in Table 2,
which is arranged as follows. Column 1: Name of the telescope where we also list
the configuration for the VLA observations. In addition to our own data, we have
also analysed some of the relevant archival VLA data on this galaxy, which are 
listed separately in Table 2.  Column 2: The frequency of the observations
where \hbox{H\,{\sc i}} denotes spectral line observations.
Columns 3 and 4: Dates of the observations and
the time, t, spent on the source in minutes. For the archival data the dates have been
taken from the image headers. However, for the observations in 2002, the precise dates
seem inconsistent with the array configurations; hence only the year is listed. The project
codes for the archival data are indicated below the table. 
Columns 5 and 6: The phase calibrator used and its flux density estimated from our 
observations. 

\begin{table}
  \caption{Observation log.}     
  \begin{tabular}{l r c r c c}
\hline
 Telescope    & Freq.             & Obs.       &      t         & Phase           & S$_{\rm cal.}$   \\
              & MHz               & date       &    min         & Calib.          &            Jy   \\
\hline
GMRT          &598                & 2004Jan04  &    180         & 2023+544         &    0.80           \\
VLA-A         &1400               & 2003Aug12  &    27          & 2023+544        &   1.11          \\
GMRT      &1408,\hbox{H\,{\sc i}} & 2004Jan26  &   240          &2023+544         &   1.03          \\
VLA-A         &4860               & 2003Aug12   &  50            & 2023+544        &   1.08          \\
VLA-A         &8460               & 2003Aug12  &    60          &2023+544         &   0.99         \\
VLA-A         &14965              & 2003Aug12   & 10         & 2023+544        & 0.87            \\
\hline
Archival \\
 
VLA-B$^1$     &4985               & 1985May25   &    45          & 1927+612          &   0.55         \\
VLA-C$^2$     &8435               & 1999July26  &    16          & 1944+548          &   0.66         \\
VLA-B$^3$     &8460               & 2002        &    9           & 1944+548          &   0.65         \\
VLA-C$^3$     &14964              & 2002        &    16          & 1824+568          &   1.68          \\
\hline
\end{tabular}
Programme code: $^1$: VH17; $^2$: AL490; $^3$: AC624
\end{table}

\begin{figure*}
\hbox{
  \psfig{file=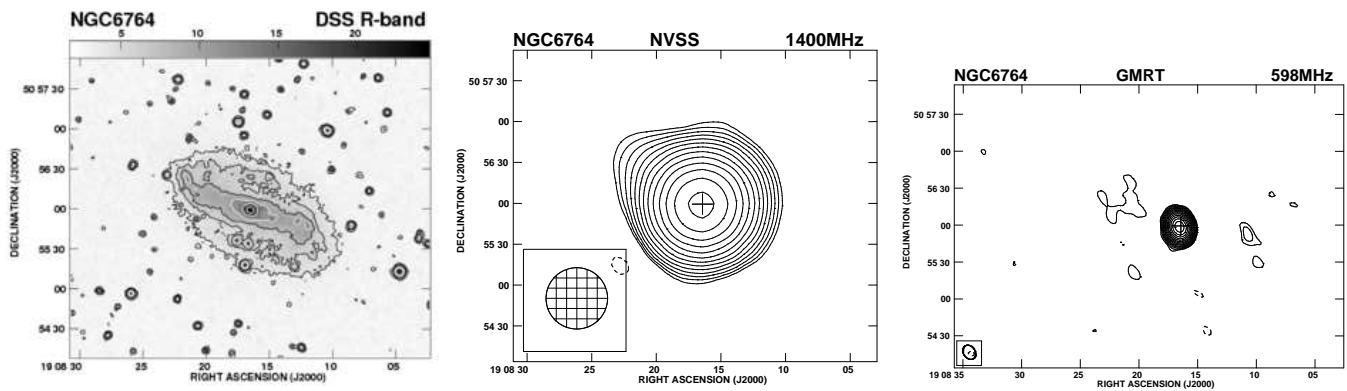,width=2.3in,angle=-90}
  \psfig{file=NGC6764NV.L.PS,width=2.3in,angle=-90}
  \psfig{file=N6764.610.STAR.1.PS,width=2.4in,angle=-90}
   }
\caption[]{Left panel: An optical R-band image of NGC 6764 from the Digitized Sky Survey (DSS). 
                       Contour levels are in arbitrary units. 
           Middle panel: NVSS image of the same region with an angular resolution of 45 arcsec.
                       Contour levels are 0.50$\times$($-$4, $-$2.82, 2.82, 4, 5.65, 8 $\ldots$ ) mJy/beam. 
                       All contour levels are in steps of $\sqrt2$, unless otherwise specified.
           Right panel: GMRT 598-MHz image of the same region of the sky with an angular
                        resolution of $\sim$11 arcsec.
                       Contour levels are 0.40$\times$($-$4, $-$2.82, 2.82, 4, 5.65, 8 $\ldots$ ) mJy/beam. 
                       The cross denotes the position of the optical nucleus from Clements (1981)
                       in all the images except Fig. 11, where the radio nucleus is marked.
}
\end{figure*}

The observations with the GMRT, which is described on the website 
{\tt http://www.gmrt.ncra.tifr.res.in}, as well as with the VLA were made
in the standard fashion, with each source observation interspersed
with observations of the phase calibrator.  The primary flux
density calibrator was 3C286 whose flux density was estimated on the
Baars et al. (1977) scale, using the 1999.2 VLA values. 
The bandwidth of the continuum observations with the GMRT at 598 MHz was 16 MHz,
while for the L-band observations it was 8 MHz. The bandwidth for all the
VLA observations was 50 MHz. The VLA data from both the IFs were combined to
make the final images.  The data analysis was done using the
Astronomical Image Processing System (AIPS) of the National Radio
Astronomy Observatory. Since GMRT data is acquired in the spectral-line mode 
with 128 spectral channels, gain and bandpass solutions were applied to each 
channel before combining them. 

The analysis of the \hbox{H\,{\sc i}} observations
was also done in the standard way. 3C286 was the primary flux density and
bandpass calibrator. The total bandwidth for \hbox{H\,{\sc i}} observations
was 8 MHz and the spectral resolution was 62.5 kHz, which corresponds to
13.5 km s$^{-1}$ in the centre of the band.  We discarded a few antennas with more than 3$\%$
fluctuations in the bandpass gains during the observations. One channel in the
beginning and eight channels towards
the end were also not included in the analysis. The AIPS task UVLIN was used
for continuum subtraction and the multi-channel data were then CLEANed using IMAGR
with the robustness parameter equal to zero.
To estimate the overall properties of the galaxy, we made images by tapering the 
data to different uv distances. 

\begin{figure}
\hbox{
  \psfig{file=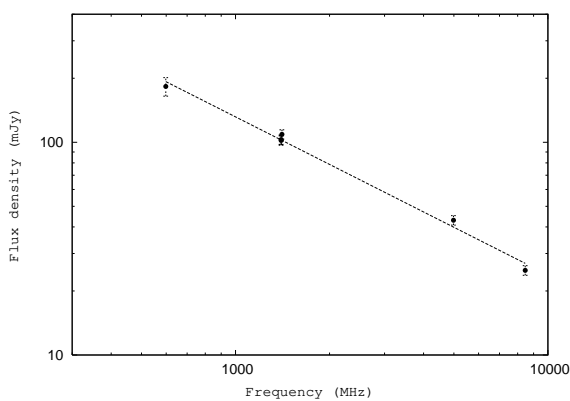,width=3in,angle=-90}
   }
\caption[]{The radio spectrum of the central region of NGC 6764. The
dashed line represents the linear least-squares fit to the data points. }
\end{figure}

We have made self-calibrated images for all the different data sets. The GMRT images 
and the archival VLA B- and C-array 8460-MHz images showed significant improvement 
after self calibration. For these data sets two cycles of phase and one cycle of 
amplitude self calibration were applied.  For the remaining 
data sets, we have presented the un-selfcalibrated images.
Some of the observed parameters of the GMRT and the VLA continuum images are presented
in Table 3 which is arranged as follows. Columns 1 and 2 are similar to that
of Table 2, except that we also list the NRAO VLA
Sky Survey (NVSS). Columns 3 to 5: The resolution of the image with the major and
minor axes being listed in arcsec and the position angle (PA) in degrees.
Column 6: The rms noise in units of mJy/beam.  Columns 7 and 8: The peak and
total flux densities in units of mJy/beam and mJy respectively. These have
been estimated by specifying a polygon around the source.
The total error in the flux density is approximately 5\%. 

\begin{figure*}
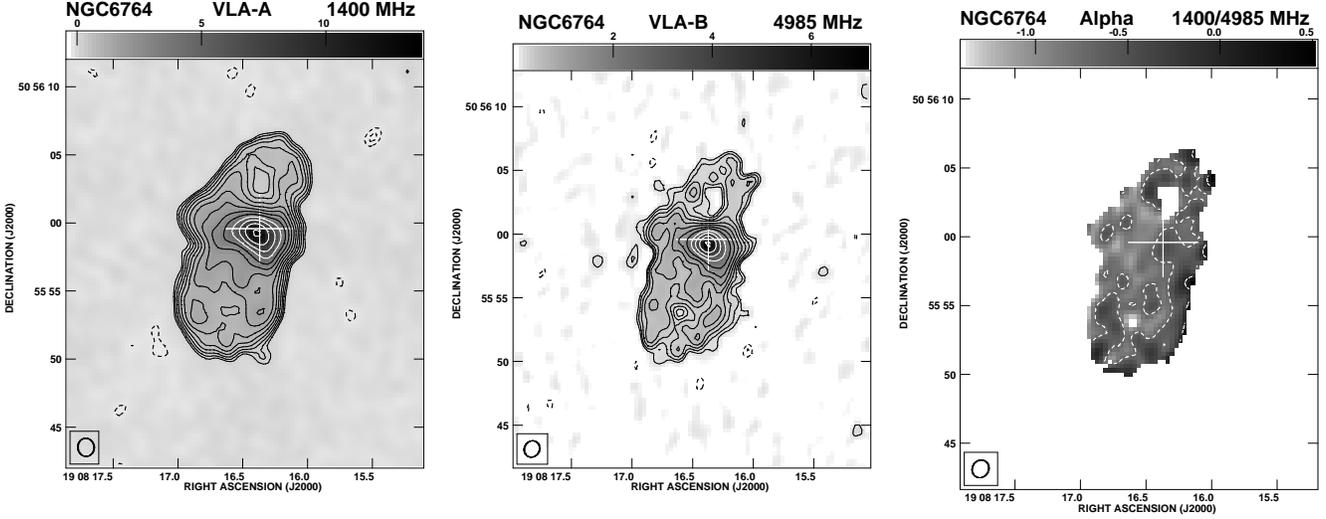

\hbox{
  \psfig{file=N6764VL.STAR.PS,width=2.3in,angle=0}
  \psfig{file=N6764AVC.STAR.PS,width=2.3in,angle=0}
  \psfig{file=6764C-L.BM.SPIX.PS,width=2.3in,angle=0}
   }
\caption[]{Left panel: VLA A-array image of the radio bubbles at 1400 MHz with an 
                       angular resolution of $\sim$1.3 arcsec. Contour levels are
                       0.052$\times$($-$4, $-$2.82, 2.82, 4, 5.65, 8 $\ldots$ ) mJy/beam in
                       steps of $\sqrt2$.
           Middle panel: VLA B-array image of the same region at 4985 MHz with an
                       angular resolution of $\sim$1.2 arcsec. Contour levels are
                       0.054$\times$($-$4, $-$2.82, 2.82, 4, 5.65, 8 $\ldots$ ) mJy/beam.
           Right panel: Spectral index map of the region obtained by smoothing the
                       4985-MHz image to the same resolution as that of the 1400-MHz image.
                       The contour level of $-$0.7 demarcates the regions of flatter and
                       steeper spectral indices.  
}
\end{figure*}

\begin{figure*}
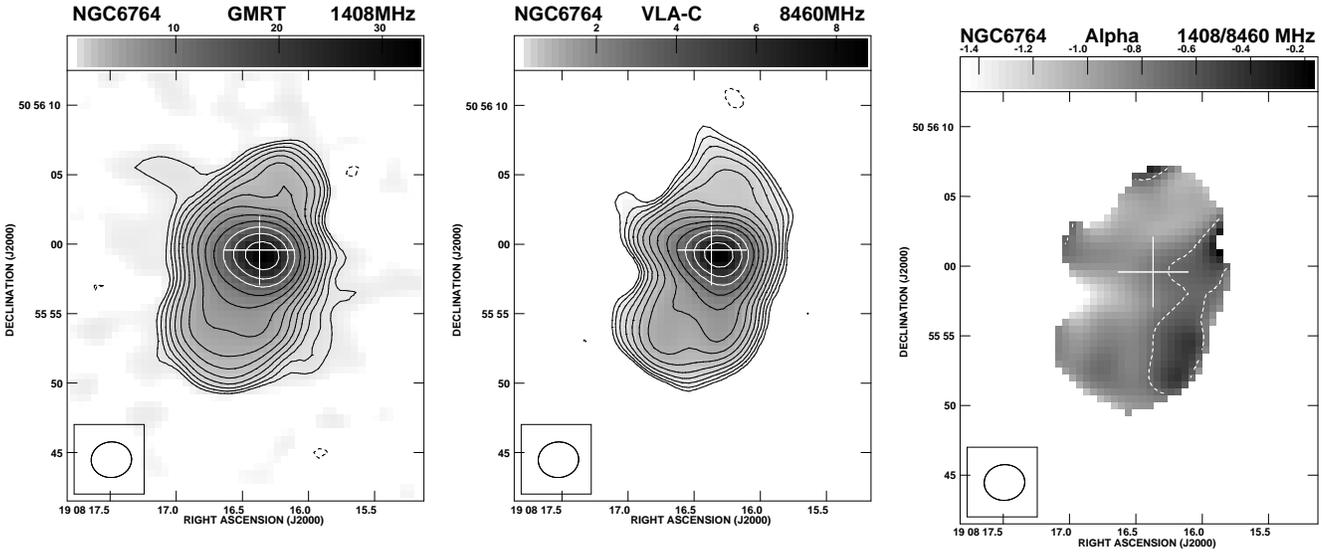

\hbox{
  \psfig{file=N6764GL.SHTT.PS,width=2.3in,angle=0}
  \psfig{file=N6764XJRG.STAR.GR.PS,width=2.3in,angle=0}
  \psfig{file=N6764X-LSH2.ST.SPIX.PS,width=2.3in,angle=0}
   }
\caption[]{Left panel: GMRT image of the radio bubbles in NGC 6764 at 1408 MHz with an
                       angular resolution of $\sim$2.7 arcsec.  Contour levels are
                       0.15$\times$($-$4, $-$2.82, 2.82, 4, 5.65, 8 $\ldots$ ) mJy/beam.
           Middle panel: VLA C-array image of this region at 8460 MHz with the same 
                        resolution as that of the GMRT image.  Contour levels are
                       0.039$\times$($-$4, $-$2.82, 2.82, 4, 5.65, 8 $\ldots$ ) mJy/beam.
           Right panel: Spectral index map of the region generated from these two images.
                       The contour level of $-$0.7 demarcates the regions of flatter and
                       steeper spectral indices.  
}
\end{figure*}
\section{Radio continuum emission}
\subsection{The radio bubbles}
In Fig. 1 we show the optical DSS R-band image of the galaxy (left
panel), the NVSS image with an angular resolution of 45 arcsec (middle panel) and the  
GMRT 598-MHz contour map with an angular resolution of $\sim$11 arcsec (right panel). 
The R-band image shows a strong stellar bar and peaks of emission which are likely to
be H{\sc ii} regions. The H$\alpha$ continuum-subtracted image of the galaxy 
(Zurita, Rozas \& Beckman 2000) shows clearly the large number of H{\sc ii} regions
and also the filamentary-like structure extending beyond the circumnuclear region,
defined here to be within $\sim$0.5 kpc from the radio nucleus.  At radio frequencies,   
the NVSS image appears somewhat extended with a deconvolved angular size of 15$\times$14 
arcsec$^2$ along a PA of 36$^\circ$, but well-within the optical extent of the galaxy. 
The GMRT 598-MHz image shows the emission to be dominated by the central radio source, 
with possible weak emission from the bar. The spectral index of the central radio source
(Fig. 2) which contains the bubbles of radio emission described below 
has been estimated from our measurements to be $-$0.74$\pm$0.02 (S $\propto$ $\nu^{\alpha}$) between 
between 598 and 8460 MHz. This shows the source to be clearly dominated by 
non-thermal emission. 

\begin{figure*}
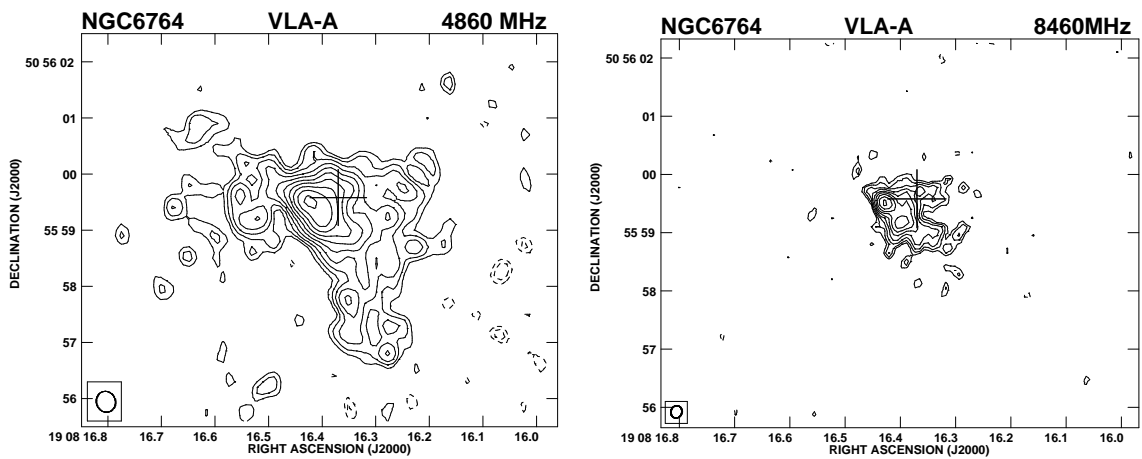

\hbox{
 \hspace{1.0cm}
  \psfig{file=N6764VC.STAR.PS,width=3.0in,angle=-90}
  \psfig{file=N6764VX.STAR.PS,width=3.0in,angle=-90}
   }
\caption[]{Left panel: VLA A-array image of the circumnuclear region at 4860 MHz with
                      an angular resolution of $\sim$0.35 arcsec.  Contour levels are
                       0.023$\times$($-$4, $-$2.82, 2.82, 4, 5.65, 8 $\ldots$ ) mJy/beam.
           Right panel: VLA A-array image of the circumnuclear region at 8460 MHz with
                      an angular resolution of $\sim$0.19 arcsec.  Contour levels are
                       0.016$\times$($-$4, $-$2.82, 2.82, 4, 5.65, 8 $\ldots$ ) mJy/beam.
   }
\end{figure*}

Published radio images of the central region of the galaxy (Wilson \& Willis 1980; 
Ulvestad, Wilson \& Sramek 1981; Condon et al. 1982; Baum et al. 1993)
have shown a `bright resolved
peak at the location of the optical nucleus and diffuse emission on 
both sides of the nucleus' (cf. Baum et al. 1993). Our images 
of the extended emission with higher
sensitivity using the VLA A-array at 1400 MHz and VLA B-array at 4985 MHz clearly show 
a bubble-like structure with a drop in brightness towards the centre of each bubble, 
located on opposite sides of the nucleus (Fig. 3: left and middle panels). These
images have been made with angular resolutions of $\sim$1.3 and 1.2 arcsec respectively.
The extents of the bubbles are $\sim$7 and 9 arcsec (1.1 and 1.5 kpc respectively) on the 
northern and southern sides respectively, while the flux densities at 1400 MHz are 42 mJy
and 62 mJy respectively. In addition to the asymmetry in the extents of
the bubbles along the minor axis of the galaxy, the radio emission in the circumnuclear region 
which is roughly orthogonal to this axis is also asymmetric relative to the radio peak 
or the position of the optical nucleus, which is marked with a $+$ sign. The emission 
on the eastern side extends for 
$\sim$5 arcsec (0.8 kpc) from the radio peak compared with 2.5 arcsec (0.4 kpc) 
on the western side.
At a somewhat lower resolution of 2.7 arcsec, our GMRT image at 1408 MHz and the VLA
archival image at 8460 MHz with the same resolution (Fig. 4: left and middle panels) show 
the bubbles to have similar asymmetric morphologies.

A spectral index image between 1400 and 4985 MHz (Fig. 3: right panel)
made by smoothing the 4985-MHz image to the resolution of the 1400-MHz image and
using only those pixels which are over 3 times the rms noise, shows that the
mean spectral index is $\sim$$-$0.72. The western side appears to have a
somewhat flatter spectral index, with a mean value of $\sim$$-$0.56, 
compared with $\sim$$-$0.77 for the rest of the source.
The spectral index image between 1408 and 8460 MHz (Fig. 4: right panel) 
shows a mean spectral index of $\sim$$-$0.85. 
This spectral index map also shows a similar trend of a 
flatter spectral index on the western side. The lateral or east-west asymmetry which is
seen in the circumnuclear region is also apparent in parts of the bubbles, as seen in the
VLA A-array image shown in Fig. 3. If this structural asymmetry, where the western side
appears closer to the peak or the axes of the bubbles, and 
spectral asymmetry are related, it could be due to an asymmetric distribution of thermal
gas. This gas could be probed by high-resolution X-ray observations. Although the ROSAT 
HRI image (Schinnerer et al. 2000) shows extended X-ray emission in addition 
to the nuclear source, the resolution is inadequate. High-resolution X-ray observations with 
Chandra are required to study the detailed distribution of this gas.

Adopting the spectral index of $-$0.74 as the mean value over the region of emission, 
and assuming the synchrotron emission to have lower and higher cut-offs at 10$^7$ and 10$^{10}$ Hz, 
respectively, a proton-to-electron energy ratio of unity, a filling factor of unity and an oblate 
spheroidal distribution for the emitting region 
(17$^{\prime\prime}$ $\times$ 9$^{\prime\prime}$ $\times$ 9$^{\prime\prime}$), 
we estimate the minimum energy density and equipartition magnetic field to be 
6.5 $\times$ 10$^{-12}$ erg cm$^{-3}$ and 8.4 $\mu$G, respectively. 
The radiative lifetime of an electron radiating in this field at 1.4 GHz is 
2.5$\times$10$^7$ yr.  These values are similar to estimates for nearby galaxies 
(e.g. Condon 1992). The 
proton-to-electron ratio is not well determined for external galaxies. For a value 
of 50 from studies of cosmic rays in our own Galaxy (e.g Webber 1991), the 
equipartition magnetic field would increase by a factor of 2.5.

\subsection{The circumnuclear region}

The full-resolution VLA-A array images of the circumnuclear region, which is
taken to be within $\sim$0.5 kpc from the radio nucleus, are presented in Fig. 5.
These images at 4860 and 8460 MHz with angular resolutions of $\sim$0.35 and 0.19 arcsec
respectively reveal the structure to be somewhat complex. 
In the lower-resolution 4860-MHz image, the emission on the western
side of the peak is more prominent and is triangular shaped, extending
for about 3 arcsec ($\sim$ 0.5 kpc) towards the south and 2 arcsec (0.3 kpc)
towards the west. The emission on the eastern side 
is more diffuse, faint and shorter in extent and contributes only 3.7 mJy of the 
total flux density of 17 mJy visible in the image. 

The higher-resolution 8460-MHz image reveals two extensions from the
radio peak, one along a PA of $-$90$^\circ$ and extending for 
$\sim$1 arcsec (0.16 kpc) and the other which is more prominent and slightly longer 
extending for $\sim$1.2 arcsec (0.2 kpc) along a PA of $-$135$^\circ$. 
Given Bridle \& Perley's (1984) definition of a radio jet, 
the south-western feature could be classified as a possible jet which 
connects the radio nucleus with more
diffuse emission which is oriented along a PA of $\sim$180$^\circ$ and 
smoothly joins on to the ridge of emission on the western side of the
southern bubble (Figs. 5). However, this feature could also be a resolved 
structure at the base of the bubble. The extensions to the east
and west could also form the base of the bubbles seen in Fig. 3. 

\subsubsection{The nuclear source}
The position of the radio peak determined from the VLA-A array 8460-MHz
image is RA: 19$^h$ 08$^m$ 16.$^s$428,  Dec 50$^\circ$ 55$^\prime$ 59.$^{\prime\prime}$47
in J2000 co-ordinates (see Table 1). This is $\sim$0.04 arcsec from the
peak in the VLA-A array 4860-MHz image (Fig. 5), but  is $\sim$0.55 arcsec from the position of the
optical nucleus (RA 19 08 16.370;  Dec: +50 55 59.58  in J2000 co-ordinates
with an uncertainty of 0.22 arcsec) determined by Clements (1981). 
This difference between the radio and optical peaks may be due to higher extinction 
close to the nucleus of this galaxy. Similar differences
have been seen in other galaxies as well, such as in  NGC1482 (Hota \& Saikia 2005).   
Although the radio peak is likely to be associated with the AGN, it is undetected 
in our VLA-A array 14940-MHz image, the 3$\sigma$ upper limit being $\sim$0.7 mJy. 
This is only marginally lower than the flux density of 0.8 mJy/beam at 8460 MHz,
suggesting that the spectrum could still be flat. A more sensitive 
VLA-A array 14940-MHz image is required to detect the possible nuclear source at
higher frequencies and determine its spectrum.  


\begin{table}
  \caption{Observed parameters of radio continuum images.}
  \begin{tabular}{lr rrr rrr}
\hline
 Telescope    & Freq.   & \multicolumn{3}{c}{Beam size}          &  rms  & S$_{\rm pk}$   & S$_{\rm tot.}$ \\
              & MHz     & maj.    & min.      & PA               &  mJy  &    mJy         &    mJy       \\
    &         & $^{\prime\prime}$ & $^{\prime\prime}$ & $^\circ$ &  /b   &   /b           &                \\
\hline
GMRT       &598      &12.4      & 9.4       & 33           &0.39       &136          &183     \\
NVSS       &1400     &45.0      &45.0       &  0           &0.50       &102          &113    \\
VLA-A      &1400     &1.33      &1.19       &7             &0.05       &14           &103       \\
GMRT       &1408     &2.89      & 2.56      &$-$85         &0.14       &33           &109   \\
VLA-A      &4860     &0.37      & 0.33      &17            &0.023      &1.6          &17        \\
VLA-A      &8460     &0.21      & 0.18      &$-$16         &0.016      &0.8          &5.9        \\
VLA-A      &14965    &0.13      & 0.11      &43            &0.221      & $<$0.7      &$<$0.7           \\
\hline
VLA-B      &4985     &1.31 &   1.18         &$-$28         &0.054      & 7.1         &43          \\
VLA-C      &8460     &2.89 &   2.56         &$-$85         &0.039      & 8.7         &25        \\
VLA-B      &8460     &1.53 &   1.29         &$-$68         &0.069      & 5.2         &20         \\
VLA-C      &14939    &2.32 &   1.06         &$-$82         &0.166      & 4.1         &14          \\

\hline
\end{tabular}\hfill\break
\end{table}

\begin{figure}
\hbox{
  \psfig{file=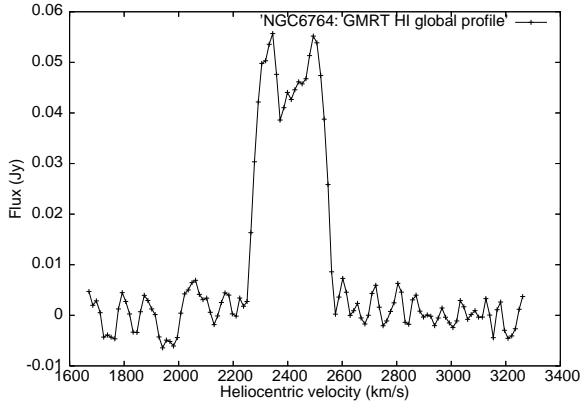,width=3.0in,angle=-90}
   }
\caption[]{GMRT  H{\sc i}-emission global profile of NGC 6764 mapped with a resolution of
37$^{\prime\prime}$.76$\times$33$^{\prime\prime}$.36  along PA $\sim$6$^\circ$ and smoothed to
27 km s$^{-1}$. 
}
\end{figure}

\begin{figure*}
\hbox{
  \psfig{file=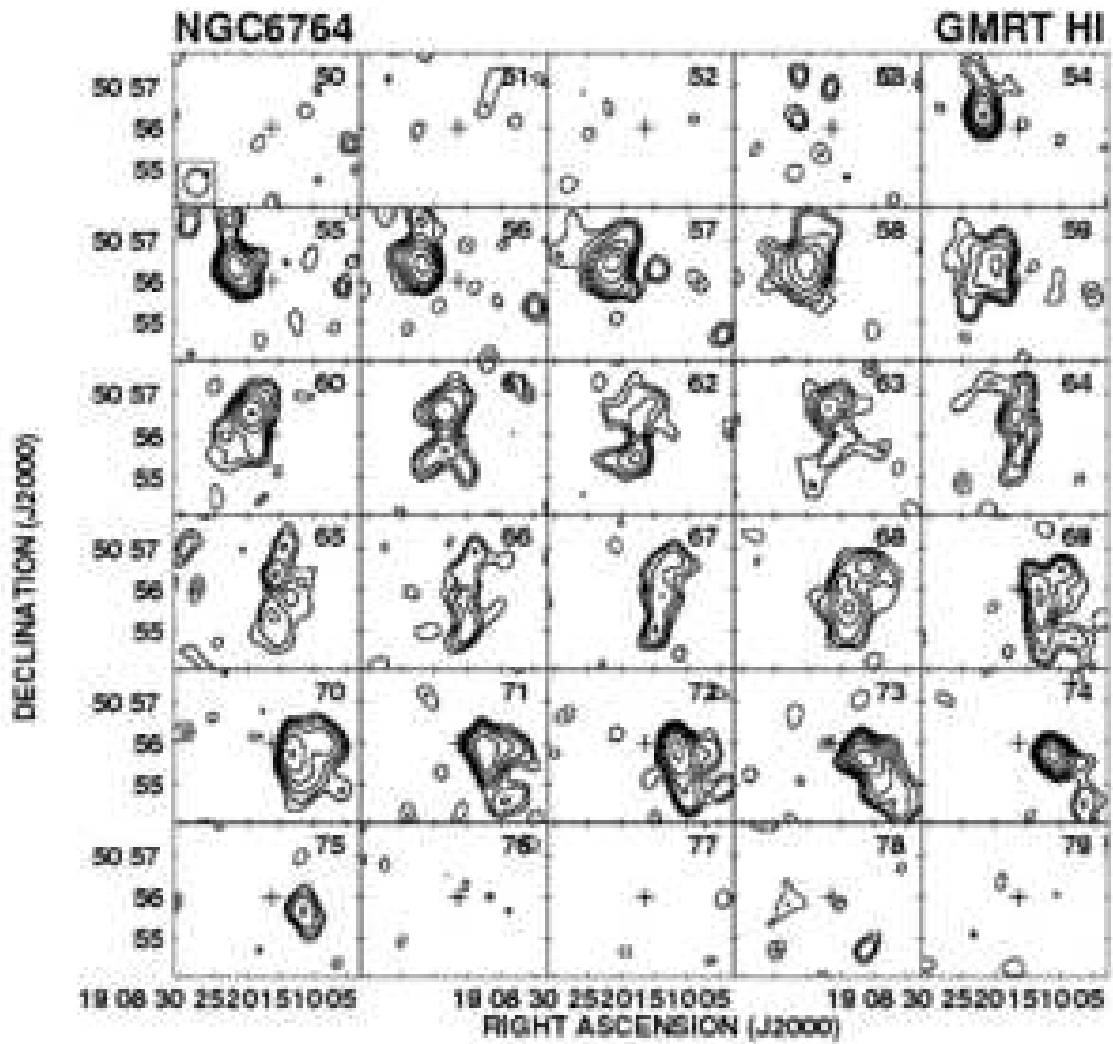,width=6.0in,angle=0}

   }
\caption[]{ NGC 6764 H{\sc i}-emission channel maps made with a resolution of
            37$^{\prime\prime}$.76$\times$33$^{\prime\prime}$.36 along PA $\sim$6$^\circ$.
            Channel 50 corresponds to a heliocentric velocity of 2602 km s$^{-1}$, while
            channel 79 corresponds to a heliocentric velocity of 2211 km s$^{-1}$. The
            systemic velocity of 2416 km s$^{-1}$ corresponds to channel 64.
            The velocity separation between adjacent channels is 13.48 km s$^{-1}$.
            The crosses in all the images denote the position of the optical nucleus.
            Peak flux: 19.8 mJy/beam. Contour levels are $-$4, $-$2.82, 2.82, 4, 
            5.65, 8 $\ldots$ mJy/beam.
}
\end{figure*}

\begin{figure*}
\hbox{
  \psfig{file=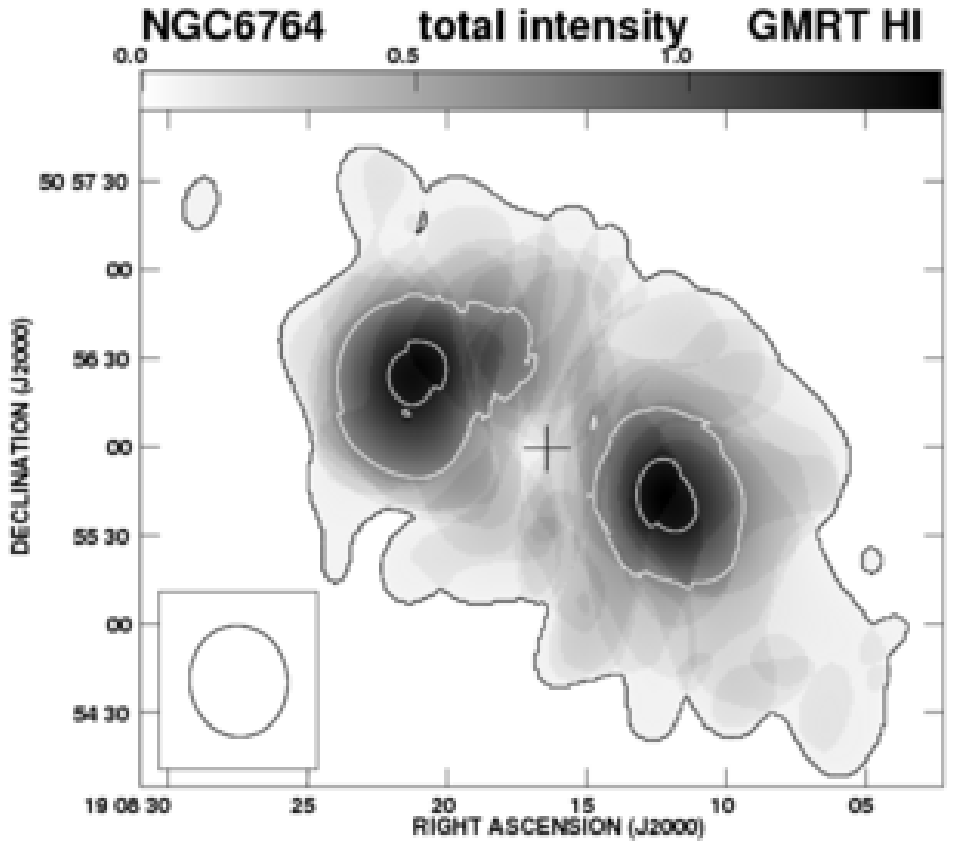,width=2.3in,angle=-90}
  \psfig{file=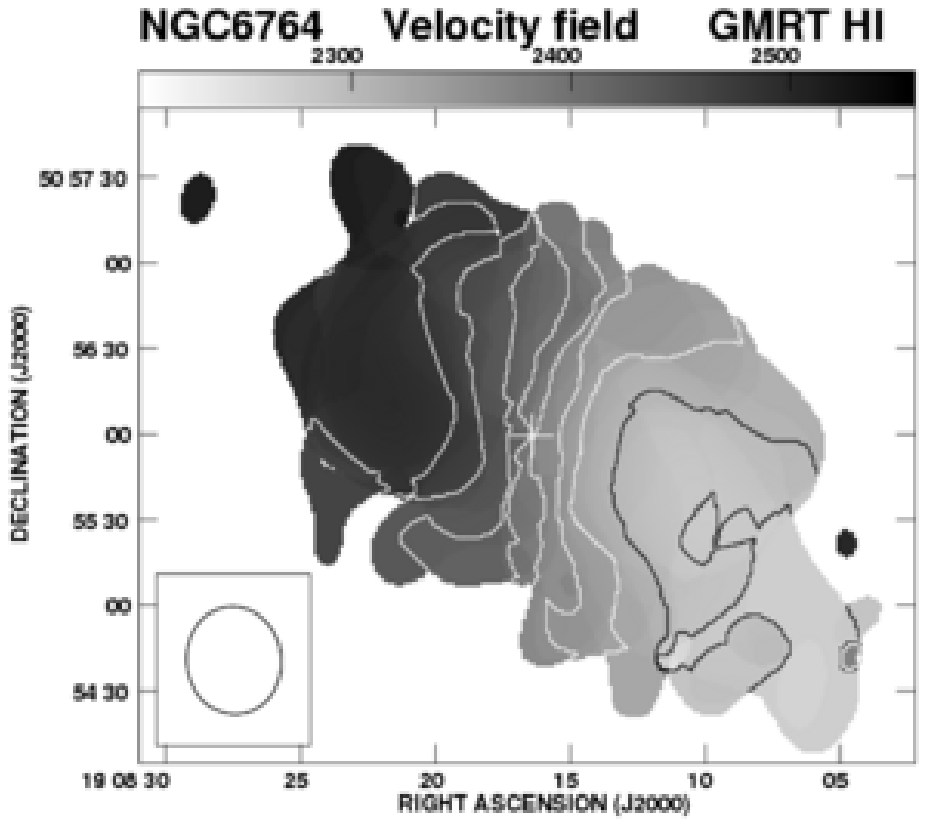,width=2.3in,angle=-90}
  \psfig{file=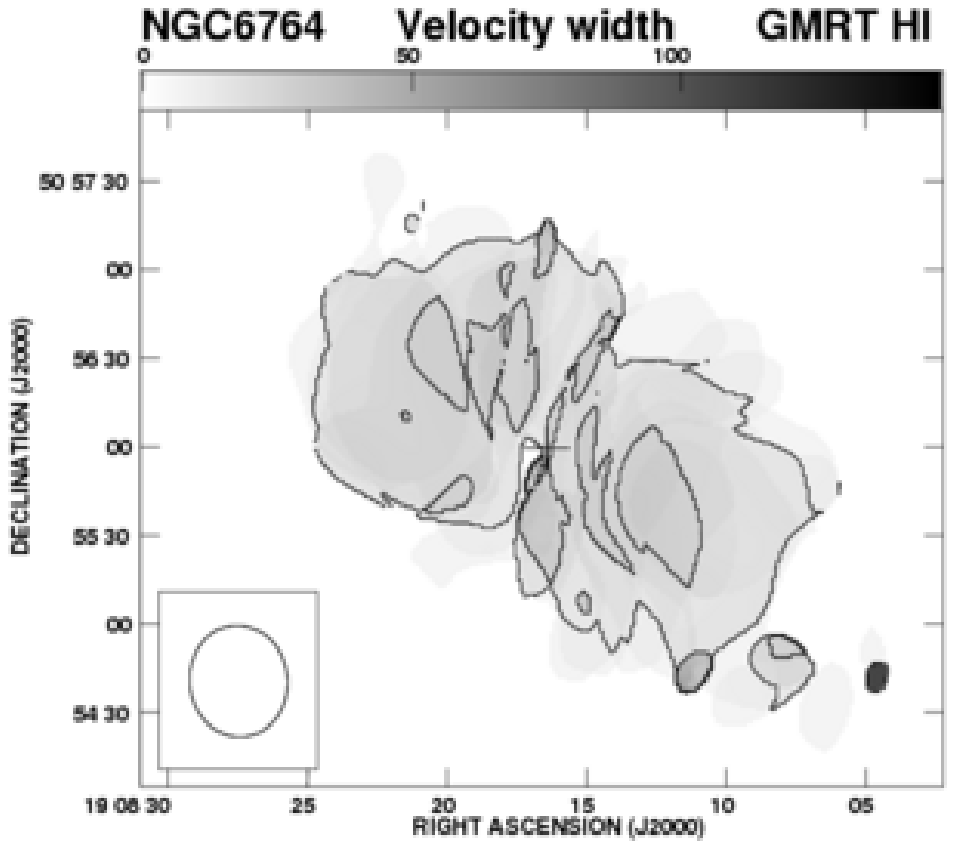,width=2.3in,angle=-90}
   }
\caption[]{ Left panel: GMRT total intensity H{\sc i} contour as well as gray scale map of
NGC 6764 made with a resolution of
37$^{\prime\prime}$.76$\times$33$^{\prime\prime}$.36 along PA$\sim$6$^\circ$. The contour levels
correspond to H{\sc i}-column densities of 5, 71 and 132 $\times$ 10$^{19}$ atoms cm$^{-2}$.
Middle panel: The H{\sc i}-velocity field (Moment-1) generated from the same data set. The
velocity contours from west to east are 2300, 2325, 2350, 2375, 2400, 2425, 2450, 2475 and 
2500 km s$^{-1}$.  The systemic velocity is 2416 km s$^{-1}$.
Right panel: H{\sc i}-velocity width map of NGC 6764 from the same data set. The contour levels
are 13 and 26 km s$^{-1}$}.
\end{figure*}


\begin{figure*}
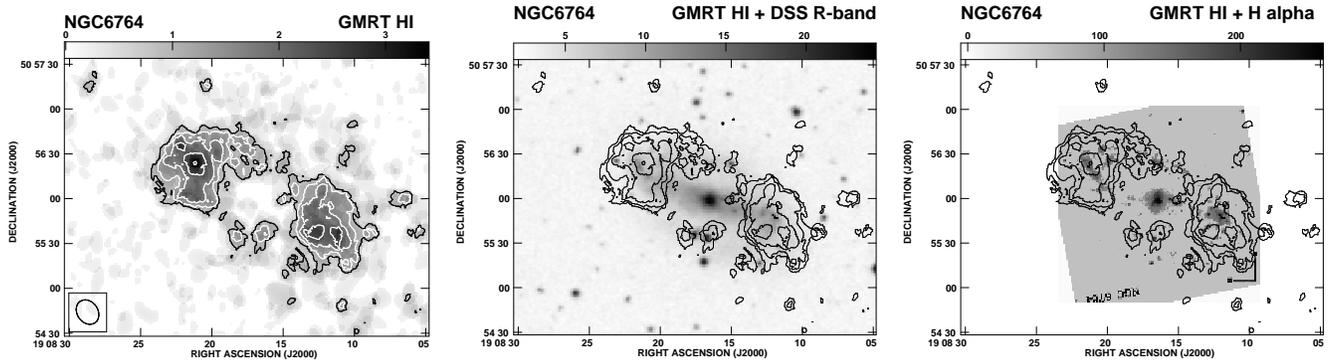

\hbox{
  \psfig{file=N6764SCT10.MOM0.PS,width=2.3in,angle=-90}
  \psfig{file=N6764DSSRHI.PS,width=2.3in,angle=-90}
  \psfig{file=N6764HAHI.PS,width=2.3in,angle=-90}
   }
\caption[]{ Left panel: H{\sc i} contour map and gray-scale image of NGC 6764 made with a resolution
of 17$^{\prime\prime}$.37$\times$13$^{\prime\prime}$.77 along PA $\sim$34$^\circ$.
The contour levels correspond to 1.21$\times$(8, 11.31, 16, 22.62, 32..) in units of 10$^{20}$ atoms cm$^{-2}$.
Middle panel: The H{\sc i} contour map with the same resolution superimposed on the DSS Red-band image.
Right panel: The H{\sc i} contour map superimposed on the H$\alpha$ image of Zurita et al. (2000).
}
\end{figure*}

\section{H{\sc i} observations}
H{\sc i} observations of NGC6764 have been reported earlier with an angular
resolution of $\sim$54 arcsec (8.8 kpc) using the VLA-D array (Wilcots, Turnbull \&
Brinks 2001). We observed this source with the GMRT with higher resolution
to (i) identify any kinematic effects of the AGN and starburst activity on the circumnuclear 
H{\sc i} gas via H{\sc i} 21-cm absorption lines towards the central continuum source, 
(ii) study the distribution of H{\sc i} gas relative to the H{\sc ii} regions,
and (iii) further investigate the nature of the observed depression in H{\sc i}
column density towards the centre of the galaxy. Wilcots et al. noted this
depression but with the relatively coarse resolution of their observations were 
unable to test the proposition that this might be due to absorption against 
the central continuum source.  The results of our observations, which have an rms 
noise of 0.6 mJy beam$^{-1}$ per channel in the full-resolution images, are described below.

\subsection{H{\sc i} emission}
In Fig. 6, we show
the global profile obtained by tapering the data to 4 k$\lambda$ and making channel
maps which have an rms noise of $\sim$1 mJy/beam and an angular resolution of $\sim$35 arcsec
(Fig. 7). In the channel maps, we confirm the extensions to the north (channel=55; 2515 km s$^{-1}$)
and south (channel=74; 2258 km s$^{-1}$) noted by Wilcots et al., which are clearly seen in our
data.  The total H{\sc i} mass calculated from the global profile 
over the velocity range 2225 to 2562 km s$^{-1}$ is 3.6 $\times$ 10$^9$ M$_\odot$.
The total H{\sc i} mass detected by Wilcots et al. (2001) is
(5.9$\pm$1)$\times$10$^9$ M$_\odot$ (revised for a distance of 34 Mpc), which implies that in our 
higher-resolution images we have resolved out some of the diffuse extended H{\sc i}-gas 
of the galaxy. 
Moment maps were generated by blanking each channel of this data cube at the 5$\sigma$ 
level. The H{\sc i} column density image and the velocity
field are shown in Fig. 8.
The velocity field  is similar to that published by Wilcots et al. (2001). The iso-velocity
contours show evidence of non-circular motions due to the existence of a bar in NGC6764.
We also see the velocity field being slightly distorted on the north-eastern and the
south-western edges of the disk, as seen by Wilcots et al. (2001). The Moment-2 map shows
the velocity width of the gas being slightly higher in the inner region
than that in the outer region (Fig. 8).

To study the H{\sc i} properties with higher resolution, we generated moment maps by
tapering the data to 10 k$\lambda$ and blanking each channel of this data cube at the 
5$\sigma$ level. The total H{\sc i} column density image with an angular resolution of
$\sim$15 arcsec shows the structure of the two well-resolved blobs of emission 
which are located on opposite sides of the circumnuclear region (Fig. 9:
left panel). A superposition of the H{\sc i} column density image on the DSS R-band image
of the optical galaxy (Fig. 9: middle panel) shows that the H{\sc i} line-emitting gas
is located in the outer regions of the galaxy, and there appears to be no 
significant H{\sc i} emission towards the centre of the galaxy which harbours the young 
starburst and the AGN.  This is unlikely to be due to absorption against the central
continuum source, which extends for only $\sim$15 arcsec, similar to the resolution of these
images, and contributes almost the entire continuum flux density seen in the NVSS image. 
The depression
in the H{\sc i} column density in the central region extends over $\gapp$50 arcsec.

A comparison of the H{\sc i} column-density image with the H$\alpha$ image of the galaxy
(Zurita et al. 2000) shows that the most prominent region in H$\alpha$ is the central
starburst source which has no significant H{\sc i} gas. Although close to the circumnuclear
region this could be due to ionisation by the central starburst and AGN,  
the depletion of H{\sc i} at larger distances could be due to an ISM phase
transition from atomic to molecular hydrogen. As mentioned earlier, there is a
concentration of molecular gas extending over $\sim$2.3 kpc along the bar of the galaxy
(Eckart et al. 1991, 1996; Kohno et al. 2001; Leon et al. 2003).

The next most prominent regions of
H$\alpha$ emission are at the ends of the bar, with weaker emission along the bar and
spiral arms. The peaks of H{\sc i} emission in the south-western blob appear displaced
from the H$\alpha$ peaks while in the northern lobe they appear to be roughly coincident
at this resolution (Fig. 9; right panel). 


\subsection{H{\sc i} absorption}

In Fig. 10, we present the H{\sc i} absorption profile towards the peak of the 
central continuum source (Fig. 4; left panel) with an angular resolution 
of  $\sim$2.7 arcsec and an rms noise of 0.6 mJy beam$^{-1}$. 
No absorption is observed towards any other part of the radio source.
The peak absorption occurs at a heliocentric velocity of 2426 km s$^{-1}$, which 
is consistent with the heliocentric systemic velocity 2416 km s$^{-1}$. In addition there 
is a weak blue-shifted component at a heliocentric velocity of $\sim$2300 km s$^{-1}$, which
corresponds to a blue shift of $\sim$120 km s$^{-1}$ relative to the systemic velocity. 
Although this feature is very weak ($\sim$2$\sigma$) and requires confirmation from 
more sensitive 
observations, it is likely to be real since it is seen in both the Stokes RR and 
LL data sets and it extends over a few channels. It is interesting to note that 
mm-wavelength observations of CO 
have shown molecular gas in emission towards the circumnuclear region which is blue 
shifted by $\sim$140 km s$^{-1}$ relative to the systemic velocity (Leon et al. 2003). 
 
The peak optical depth, calculated using the background peak continuum flux density 
of 33 mJy from the area of 9 arcsec$^2$ against the peak is 0.06. 
The total optical depth, calculated with data points over the full width at zero intensity 
(2184$-$2548 km s$^{-1}$) is 0.5. Assuming a spin temperature of 100 K this corresponds
to a total N$_{H{\sc i}}$ $\sim$ 1.23 $\times$ 10$^{21}$ cm$^{-2}$. The total mass of the 
absorbing H{\sc i} clouds is 2.35 $\times$ 10$^{6}$M$_\odot$. The blue-shifted component
spread over 2184$-$2346 km s$^{-1}$ has an optical depth of 0.13 which corresponds
to an N$_{H{\sc i}}$ $\sim$ 3.1 $\times$ 10$^{20}$ cm$^{-2}$. Although the blue shift
could be due to non-circular motions due to the barred potential, such shifts may also
be caused by gas clouds driven outwards by the circumnuclear starburst or AGN.
In the latter scenario, the outflowing ($\sim$120 km s$^{-1}$) H{\sc i} cloud, which has 
an estimated mass of 5.9$\times$10$^{5}$M$_\odot$,
assuming the cloud to have an area of 9 arcsec$^2$, has a
kinetic energy of 8.5$\times$10$^{52}$ergs.

\section{Discussion}

\begin{center}
 \begin{table*}
 \centering
 \begin{minipage}{170mm}
  \caption{Comparision with other radio-bubble galaxies.}
 \hspace{0.5cm}  \begin{tabular}{@{}cccccccccc@{}}
\hline
Galaxy   &Morph. & AGN          & Dist.    & Size  &  Size   &  S$_{\rm b+n}$(1.4) & L$_{1.4}$ (10$^{20}$) &log(L$_{\rm FIR}/L_\odot$) & Refs.\\
 name     &class& type      & Mpc & $^{\prime\prime}$ &  kpc &    mJy          &  W/Hz  &   &\\
 (1)      & (2) &  (3)      & (4) &        (5)        & (6)  &   (7)           &  (8)   &  (9)  &  (10)  \\
\hline
NGC1068   &SAb& Sy1,Sy2          & 16.0    & 14    & 1.0      & 4221            &    1296      &10.98    &1,2\\
NGC2782   &SABa& Sy1,SB          & 36.0    & 14    & 2.4      & 107             &     166      &10.44    &3,4\\\
NGC2992   &SA,pec& Sy1,Sy2       & 32.5    & 8     & 1.2      & 168             &     213      &10.26    &5,6 \\
NGC3079   &SBc& Sy2,L            & 15.7    & 45    & 3.4      & 409             &     121      &10.53    &7,8\\\
NGC3367   &SBc& Sy,L             & 42.7    & 75    &15.3      & 51              &     112      &10.49    &9,10\\\
NGC4051   &SABbc& Sy1.5          & 9.9     & 10    & 0.4      & 20              &       2.3    & 9.47    &11,12\\\
M51 &SAbc,pec&H{\sc ii},Sy2.5,L  & 8.4     & 15    & 0.6      & 45              &       3.8    &10.38    &13,14\\\
NGC5548   &SA0/a& Sy1.5          & 72.3    & 10    & 3.4      & 23              &     147      & 9.98    &12,15\\\
NGC6764   &SBbc& Sy2,L           & 34.0    & 16    & 2.6      & 106             &     147      &10.31    &16,17\\\
Circinus  &SAb& Sy2              & 6.1     & 270   & 8.1      & -               &    -         &$\sim$10.45&18,19\\
\hline
\end{tabular}\hfill\break
1 Cecil, Bland \& Tully 1990; 2 Pogge 1988; 3 Jogee, Kenney \& Smith 1998; 4 Yoshida, Taniguchi \& Murayama 1999; 
5 Chapman et al. 2000; 6 Veilleux, Shopbell \& Miller 2001; 7 Cecil et al. 2001; 8 Irwin \& Saikia 2003; 
9 Garc\'{i}a-Barreto, Franco \& Rudnick 2002; 10 Garc\'{i}a-Barreto et al. 2005;
11 Christopoulou et al. 1997; 12 Baum et al. 1993; 13 Cecil 1988; 14 Ford  et al. 1985; 15  Wilson et al. 1989; 
16 This paper; 17 Zurita, Rozas \& Beckman 2000; 18 Elmouttie et al. 1998a; 19 Elmouttie et al. 1998b. \\
\end{minipage}
\end{table*}
\end{center}

\begin{figure}
\hbox{
  \psfig{file=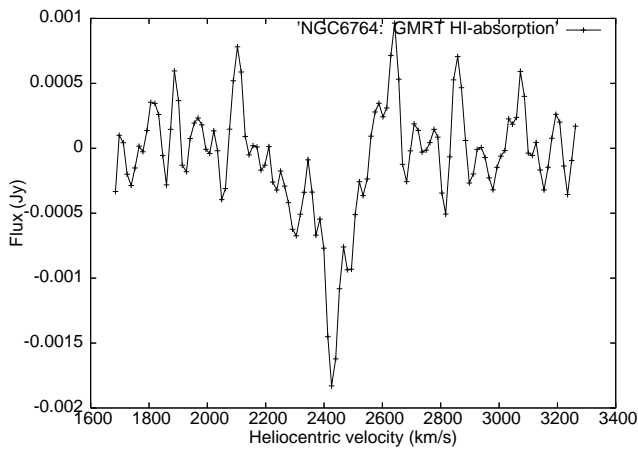,width=3.3in,angle=-90}
   }
\caption[]{
GMRT  H{\sc i}-absorption spectrum taken from a 3$^{\prime\prime}$$\times$3$^{\prime\prime}$
region around the radio continuum peak of NGC 6764. The spatial resolution of the data cube is
2$^{\prime\prime}$.90$\times$2$^{\prime\prime}$.55 along PA $\sim$-87$^\circ$; the spectrum
has been smoothed to 27 km s$^{-1}$.
}
\end{figure}

\subsection{Radio bubbles}
\begin{figure}
\vbox{
 \hspace{1.0cm}
 \psfig{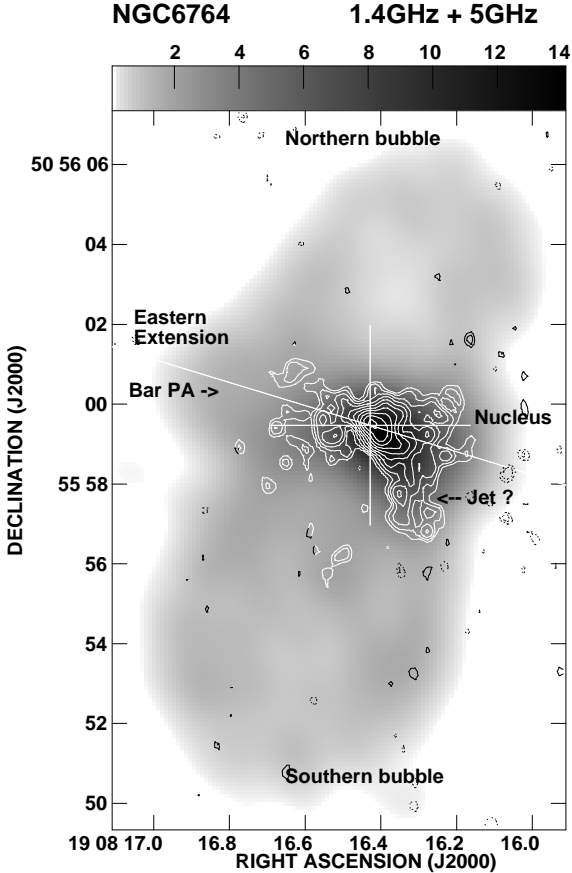}
   }
\caption[]{Contour map of the radio-continuum emission from the circumnuclear region
of NGC6764 at 4860 MHz with the VLA A-array (Fig. 5, left panel) is superimposed on the gray scale
image of the radio bubbles at 1400 MHz with the VLA A-array (Fig. 3, left panel). The bubbles, the
eastern extension and a possible jet are labelled. The cross marks the position of 
the radio nucleus.  The line at a PA of 73$^\circ$ shows the orientation of the stellar bar.
}
\end{figure}

At radio continuum wavelengths, the most striking features are the bubbles of
non-thermal plasma with a size of $\sim$1.1 and 1.5 kpc on the northern and
southern sides respectively of the circumnuclear region, and oriented roughly orthogonal 
to the stellar bar and the major axis of the galaxy (Fig. 11). There also appears to be an 
extension on the eastern side extending beyond the base of the bubbles. In addition, 
smaller-scale structure with a size of $\sim$0.5 kpc but asymmetric in extent 
relative to the radio nucleus is seen oriented roughly along the stellar bar.  
There is also a jet-like structure starting from the nucleus along
an initial PA of $-$135$^\circ$ which appears to connect the nucleus to more 
extended emission towards the south-west. These features are shown in Fig. 11 where 
the smaller-scale structure is shown superimposed on the larger-scale bubbles.

We first discuss the origin of these structures in this composite
galaxy which harbours both a young starburst and an AGN. It is worth noting
here that the activity in this galaxy does not appear to be triggered
by interaction with any companion galaxy, since there is no evidence of any 
nearby galaxy within a distance of $\sim$160 kpc (Wilcots et al. 2001). However,
the H{\sc i} observations do show evidence of a weak perturbation but this is unlikely
to significantly affect the starburst and AGN activity seen in this galaxy.

In order to understand the nature of these bubbles, we have compiled a representative sample
of nearby galaxies with similar structures from the literature. A bubble is defined to
be one whose width is larger than approximately half the size of its
length. This working definition has been adopted to distinguish it from jets
or jet-like features and normal lobes of radio emission formed by these jets whose 
axial ratios are usually higher than $\sim$2 (e.g. Leahy, Muxlow \& Stephens 1989). 
The sample of nearby galaxies with bubbles is listed in 
Table 4, along with some of their properties. The table is arranged as follows.
Column 1: name of the galaxy; column 2: galaxy classification from NED;
column 3: classification of the AGN from NED where Sy denotes a Seyfert galaxy, 
L a LINER galaxy, SB a starburst and H{\sc ii} a galaxy with prominent 
H{\sc ii} regions; column 4: distance to the galaxy
in Mpc; columns 5 and 6: total extent of the bubbles in arcsec and kpc 
respectively.  Columns 7 and 8: flux density and radio luminosity at 1.4 GHz
of the bubbles and any core or jet emission.  These have been estimated from the
FIRST images, except for NGC2992 for which we have used the NED image 
(Ulvestad \& Wilson 1989) and NGC6764 for which we have used the results presented in 
this paper. The flux density 
of the bubbles and any core or jet emission has been estimated by specifying a polygon around 
this region. The flux density value for the bubbles in
Circinus is not available. Column 9: The FIR luminosity of the galaxies. Except for 
Circinus, for which we have estimated the value from Elmouttie et al. (1998b), these have been
taken from Condon et al. (1990, 1998) and scaled to the distances listed in 
column 4. Column 10: References for the radio, X-ray and/or H$\alpha$ images.

It can be seen that the total extents of the bubbles range from $\sim$0.4 to 15 kpc, with
a median value of $\sim$2.5 kpc, similar to that of NGC6764. While some bubbles
lie within the scale-height of the interstellar medium (ISM) of the parent galaxy, 
others extend beyond it.
In these radio bubbles, the ionised gas seen in optical emission lines 
such as H$\alpha$ or [O{\sc ii}] tend to be spatially related to the radio emission. 
The radio luminosity of the bubbles and any core or jet emission ranges from 
$\sim$2.3$\times$10$^{20}$ 
to $\sim$1.3$\times$10$^{23}$ W Hz$^{-1}$ with a median value of $\sim$1.4$\times$10$^{22}$
W Hz$^{-1}$, while the infrared luminosity 
of the galaxies ranges from 9.47 to 10.98 in units of log(L$_{\rm FIR}$/L$_\odot$) with a 
median value of 
$\sim$ 10.3. Although the galaxies are luminous at infrared wavelengths, their luminosity is
well below the threshold for defining these as ultra-luminous infrared galaxies. 

In Table 4, it is striking that all the sources with non-thermal 
bubbles of radio emission have an AGN associated with it. It is worth
comparing this with archetypal starburst galaxies such as  M82 and NGC253,
whose infrared luminosities of 10.38 and 10.26 in units of log(L$_{\rm FIR}$/L$_\odot$) 
respectively, are comparable with the sources in the sample of galaxies with
bubbles. These two galaxies are dominated by the circumnuclear starburst with
no clear evidence of an AGN, although there have been attempts to identify one
(Wills et al. 1999; Mohan et al. 2002). High-resolution radio images of
the nuclear regions of these two galaxies reveal a large number of compact
radio components (Muxlow et al. 1994; McDonald et al. 2002 and references therein;
Collison et al. 1994; Ulvestad \& Antonucci 1997 and references therein) 
which are a mixture of H{\sc ii} regions and supernova remnants, while lower-resolution 
observations reveal a more extended
halo of emission (Seaquist \& Odegard 1991; Reuter et al. 1992; Carilli et al. 1992). 
Although outflows along the minor axes are visible in 
both these two galaxies in X-rays and H$\alpha$ (Strickland et al. 2004),
there are no bubbles of non-thermal plasma similar to those seen in our 
sample of galaxies. The archetypal starburst galaxy in the southern hemisphere,
NGC1808, which is also dominated by a starburst with no unambiguous identification
of an AGN (Forbes et al. 1992; Jim\'{e}nez-Bail\'{o}n et al. 2005), has extended
non-thermal emission with no clear bubbles of radio emission (e.g. Dahlem et al.
1990). The small-scale radio structure is dominated by a number of compact components
(Saikia et al. 1990; Collison et al. 1994).
Also, in a recent study of the superwind galaxy NGC1482,
which has a remarkable hourglass-shaped optical emission-line outflow as well 
as bipolar soft X-ray bubbles of emission (Veilleux \& Rupke 2002; Strickland et al. 
2004), there is no non-thermal radio emission or bubbles associated with the superwind 
although radio emission is seen associated with the disk of the galaxy (Hota \& Saikia 2005). 
Its infrared luminosity of log L$_{\rm FIR}$=10.66 is similar to the sources
with bubbles listed in Table 4. Although 
the high-resolution radio observations of the  nuclear region of NGC1482
show a peak of emission with a steep radio spectrum, and more diffuse
emission with secondary peaks, it is not clear if this feature could be
associated with an AGN (Hota \& Saikia 2005). It is relevant to note here 
that, from optical spectroscopic observations, Kewley et al. (2000) have 
classified it to be a starburst galaxy without an AGN. Considering these
aspects, it is tempting to speculate that the formation of bubbles is
closely linked to the existence of an AGN.  It is interesting
to note that Colbert et al. (1996) have highlighted similar differences in 
the radio structures between Seyfert and starburst galaxies. 

Thus although the AGN possibly gives rise to the bubbles of non-thermal radio 
emission, these would interact with the external environment in the host galaxy. 
A galactic wind generated by a starburst (cf. Veilleux et al. 2005) along with 
effects of buoyancy (Gull \& Northover 1973; Stone, Wilson \& Ward 1988) could lead 
to the bubbles of plasma being oriented along the minor axis of the galaxy. In this 
case, the radio jet close to the nucleus could be misaligned with the
minor axis, as seen in NGC6764 as well as in the well-studied galaxy NGC3079
(cf. Irwin \& Seaquist 1988; Kondratko et al. 2005). Using the formalism of
Stone et al. (1988), we estimate the time scale for the bubbles to reach the
observed distances due to buoyancy to be $\sim$35 Myr. This is likely to
be an upper limit because of the presence of an AGN as well as galactic wind. 
The extended X-ray emission from a region similar to that of the radio bubbles 
(Schinnerer et al. 2000) and the velocity structure in the H$\alpha$ gas (Rubin et al.
1975) suggests the presence of a galactic wind in this starburst galaxy.
Estimating an outflow velocity of $\sim$80 km s$^{-1}$ for the ionised gas
observed by Rubin et al., the dynamical time scale for the H$\alpha$
gas to reach distances of $\sim$1.5 kpc is $\sim$20 Myr. For comparison,
the circumnuclear star formation history suggests two starbursts with ages of  
3$-$5 Myr and between 15 and less than 50 Myr (Schinnerer et al. 2000),
suggesting that both episodes of star formation could have affected the 
detailed structure of the bubbles.


\subsection{Kinematic effects on the ISM}

\begin{figure}
\vbox{
  \psfig{file=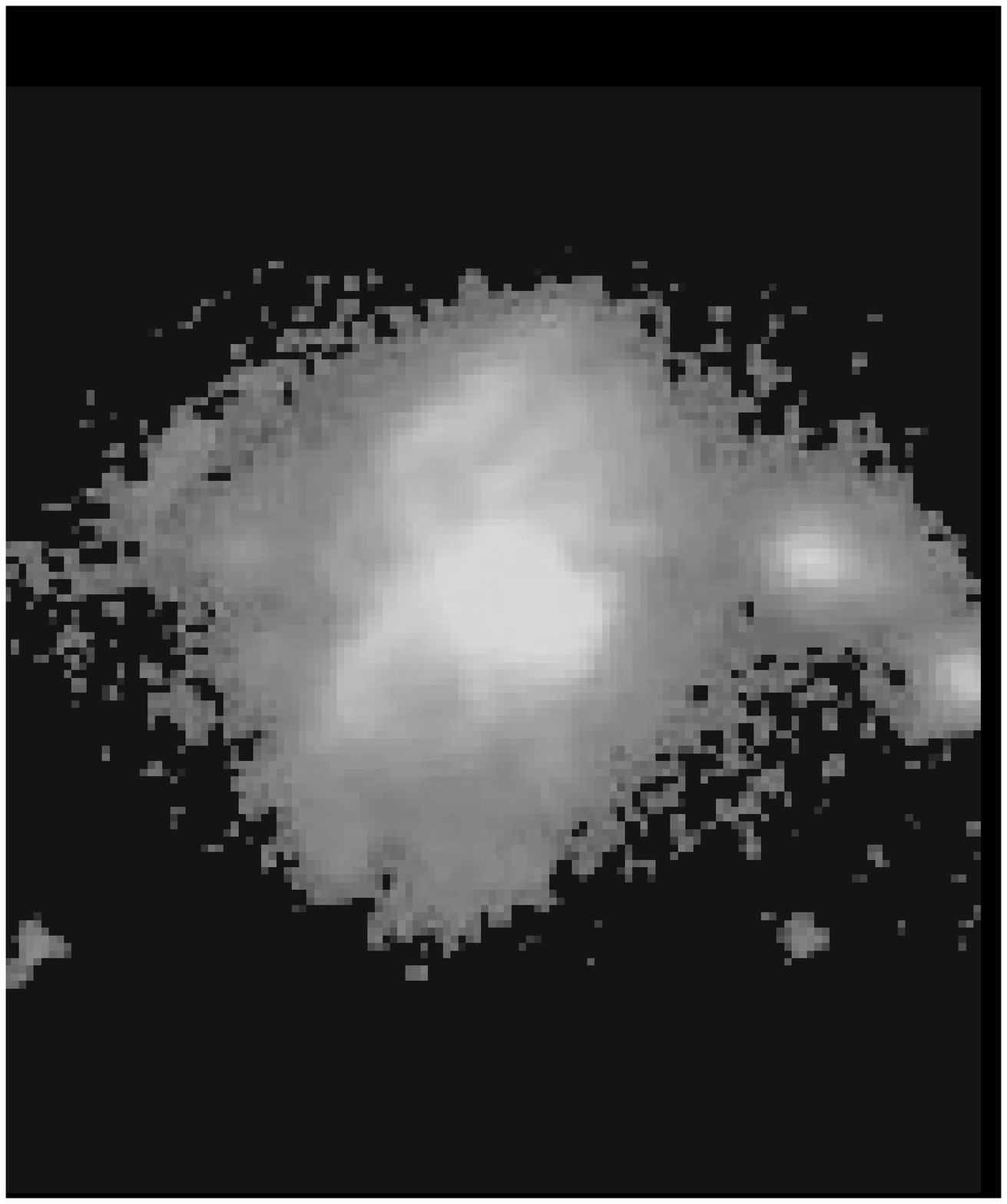,width=3.3in,angle=0}
  \psfig{file=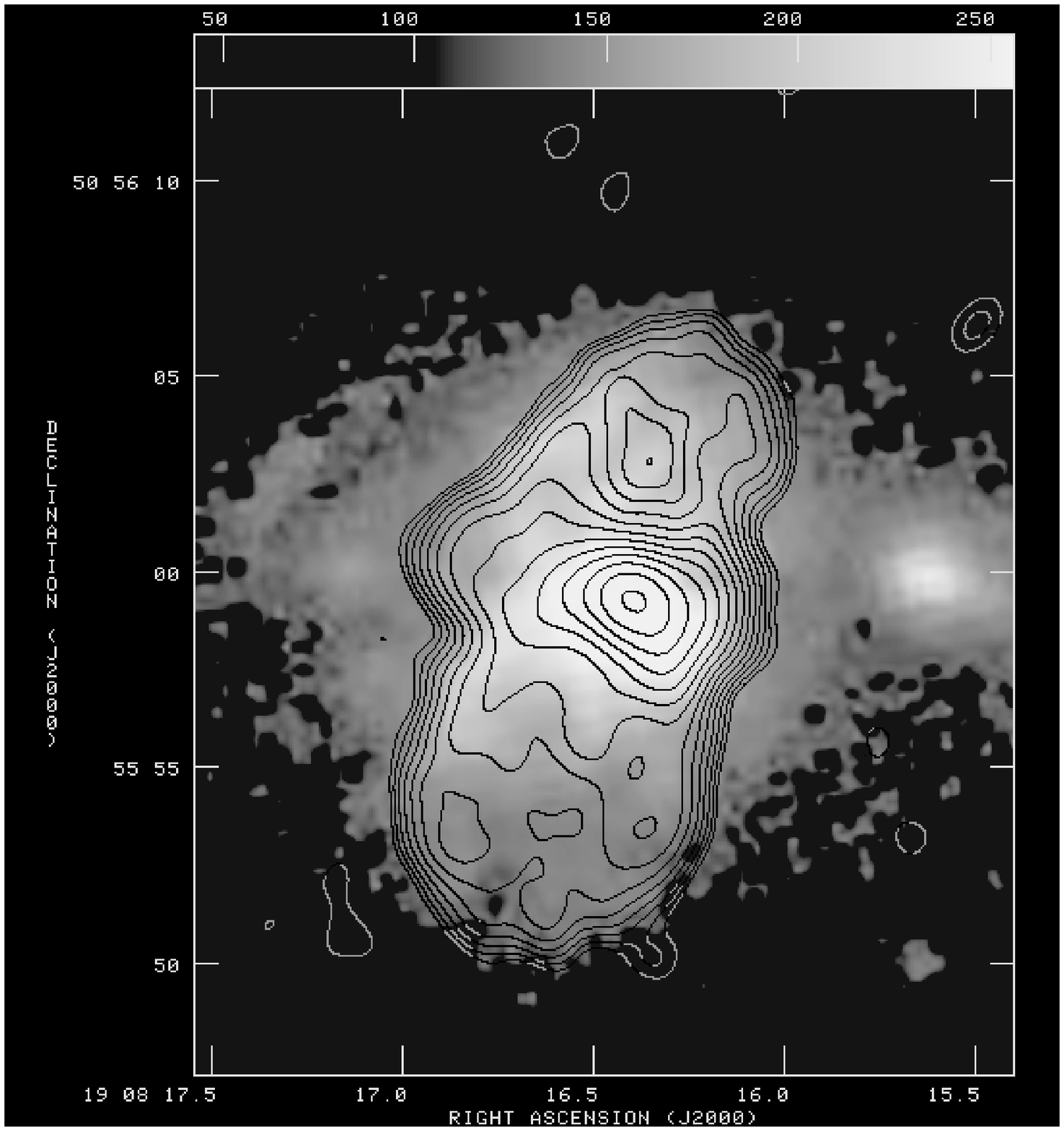,width=3.3in,angle=0}
   }
\caption[]{Top panel: H$\alpha$ image of the circumnuclear region of NGC 6764 from the published
image of Zurita et al. (2000), showing the H$\alpha$-filaments.
Bottom panel: Contour map of the radio bubble at 1400 MHz with the VLA A-array (Fig. 3, left
panel) is superimposed on the same H$\alpha$ image in gray scale.
}
\end{figure}

\begin{figure}
\vbox{
  \psfig{file=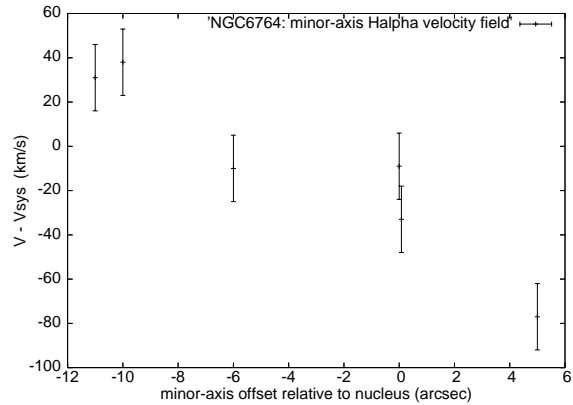,width=3in,angle=-90}
   }
\caption[]{The H$\alpha$ and [N{\sc ii}] velocity field measured along the minor-axis of the 
bubbles. Data plotted from the measurements of Rubin et al. (1975).
}
\end{figure}


Interaction of radio jets and bubbles, as well as the galactic wind with
the ambient medium could affect the kinematic properties of the ISM. 
However, in the case of NGC6764 the radio features are likely to play a
relatively smaller role. The total minimum energy in the radio features
is only $\sim$4.8$\times$10$^{54}$ ergs,
which corresponds to a pressure of 2.2$\times$10$^{-12}$ dynes cm$^{-2}$,
similar to values for the ISM in the Milky Way.
The total energy in the radio features of NGC6764 are significantly
smaller than typical values for the total energy in the outflows. 
Although the total energy in the various components in the outflow in NGC6764
is not known, the kinetic energy in the outflow of only the warm ionised and
the neutral gas components in typical
galaxies with galactic winds are $\sim$10$^{53}$ to 10$^{56}$ 
and $\sim$10$^{56}$ to 10$^{60}$ ergs respectively (Veilleux et al. 2005).

A continuum-subtracted H$\alpha$ image of the central region of the galaxy
(Zurita et al. 2000) shows a number of filamentary structures, reminescent of 
those seen in galaxies with galactic winds. We have determined
the positional information of this H$\alpha$ image by comparing the pixel positions of three
stars which are visible in the H$\alpha$ image with their positions determined
from the DSS image. A superposition of the radio continuum image at 1400 MHz with
an angular resolution of 1.2 arcsec (0.2 kpc) on the H$\alpha$ image
(Fig. 12) shows some correlation of the radio emission with the H$\alpha$ features,
which are possibly driven outwards by a galactic wind. 

Spectroscopic observations of NGC6764 by Rubin, Thonnard \& Ford (1975) showed
that along the axis of the bubbles, the ionised gas appears red-shifted on the
northern side and blue-shifted on the southern side by  up to $\sim$80 km s$^{-1}$  
relative to the systemic velocity of the galaxy (Fig. 13). These motions are likely 
to be caused by a bi-polar outflow due to the circumnuclear starburst. 

In addition to the ionised gas component, the outflow due to the 
circumnuclear starburst could also affect the molecular and atomic gas components in 
the galaxy. For example, interaction of the bubbles with the molecular gas can give 
rise to CO gas compression and also affect their kinematic properties. Such effects
have been seen in detailed studies of individual starburst galaxies. For example, in
the starburst galaxy NGC253, Sakamoto et al. (2006) find that the 
circumnuclear molecular disk harbouring the starburst is highly disturbed on small scales by 
individual young clusters and supernovae and globally by the well-known superwind.
In the map of NGC6764 published by Eckart et al. (1991), the line ratio CO(2-1)/CO(1-0)
shows higher values of $\sim$0.9 along the orientation of the bubbles but smaller
values of $\sim$0.5  
near the arms. Such higher values of the line ratio are expected when CO gas is 
compressed. Leon et al. (2003) find evidence of CO gas which is blue shifted by
$\sim$140 km s$^{-1}$ and have a very high line ratio of $\sim$2
for CO(2-1)/CO(1-0). As discussed earlier (Section 4.2) our H{\sc i} absorption 
spectrum shows a possible feature at a blue-shifted velocity of $\sim$120 km s$^{-1}$, 
which is similar to that of the molecular gas observed in CO.  
A similar blue-shifted feature 
was also noted in the superwind galaxy NGC1482 by Hota \& Saikia (2005), which was
also interpreted to be due to the circumnuclear starburst.

Hydrodynamical simulations of starburst-driven winds show that the cool gas
could expand laterally in the disc of the galaxy, be carried vertically outwards
by the tenuous superwind or be entrained in the interface between the hot, superwind
fluid and the cool, dense ISM (e.g. Heckman et al. 1990; Heckman et al. 2000).
The kinematic and physical properties of the different components discussed above
are broadly consistent with the results of these simulations of winds.

\section{Summary and concluding remarks}

We have presented radio continuum and H{\sc i} observations of NGC6764,
which has a young starburst as well as a Seyfert nucleus. The results
are briefly summarised here.
 
\begin{enumerate}

\item The low-resolution GMRT image at 598 MHz with an angular resolution of 
$\sim$11 arcsec (1.8 kpc)  shows that almost all the radio emission 
from the galaxy comes from the central region of the source, with very little 
emission from the disk of the galaxy. The VLA A-array image at 1400 MHz with
an angular resolution of $\sim$1.3 arcsec accounts for more than 90 per cent 
of the flux density visible in the NVSS image. This central source has a 
non-thermal spectrum with a spectral index of $-$0.74$\pm$0.02 between 
$\sim$600 MHz and 8 GHz.

\item With an intermediate resolution of 1.3 arcsec (0.2 kpc),
the central source is resolved into two bubbles of non-thermal radio 
emission which are shell-like with a depression in the centre. The
ridges of emission appear to overlap with H$\alpha$ filaments seen
in a continuum-subtracted H$\alpha$ image of the galaxy. The bubbles are 
oriented along the minor axis of the galaxy with a total extent of
$\sim$2.6 kpc. The western edge of the bubbles tends to have a flatter
spectral index than the rest of the bubbles. 

\item The image at 4860 MHz with a resolution of $\sim$0.35 arcsec (0.05 kpc)
shows emission along the major axis of the galaxy as well as extensions which
appear to connect to the ridges of emission in the bubbles.

\item Our highest-resolution image (0.19 arcsec; 0.03 kpc) reveals a compact
source, which is possibly associated with the nucleus of the galaxy, and
is slightly offset ($\sim$0.55 arcsec) from the position of the optical nucleus 
determined by Clements (1981).  In addition there is a possible radio jet 
along a PA of $-$135$^\circ$ which appears to connect to more extended
emission towards the south.

\item  A compilation of a representative sample of ten galaxies with non-thermal 
bubbles shows that these are all associated with an AGN. However, 
the detailed radio and optical structures, as well as the distribution of
atomic and molecular gas,  could be affected by a galactic
wind due to a starburst in addition to any AGN activity.
Prominent starburst galaxies such as M82, NGC253, NGC1482 and NGC1808
have similar infrared 
luminosity, but do not exhibit such bubbles of radio emission, suggesting that 
these are not usually caused by the circumnuclear starburst. 

\item  The \hbox{H\,{\sc i}} observations, which are of higher resolution than
those reported earlier by Wilcots et al. (2001), show that the two main 
peaks of emission are roughly coincident with the ends of the bar. 
There is a depletion of H{\sc i} towards the central region. Although this 
is likely to be due to the circumnuclear starburst and AGN close  to  the
nucleus, at larger distances it could be due to phase transition in the ISM from
atomic to molecular hydrogen. 

\item  The peak absorption feature in the H{\sc i}-absorption profile 
has an optical depth of $\sim$0.06 and a  heliocentric velocity of 2426 km s$^{-1}$, 
consistent with the systemic velocity of the galaxy. The \hbox{H\,{\sc i}} mass 
estimated from the absorption profile is $\sim$2.4$\times$10$^6$ M$_\odot$. 
There is a suggestion of a weak absorption feature at a blue-shifted 
velocity of 120  km s$^{-1}$, which requires confirmation from more 
sensitive observations. A feature with a similar blue-shifted velocity has also 
been reported from observations of CO in emission, suggesting
that the circumnuclear starburst and nuclear activity may have affected the 
kinematics of the atomic and
molecular gas components of the ISM in the host galaxy, in addition to the
ionised gas seen in H$\alpha$ and [N{\sc ii}].

\end{enumerate}

\section*{Acknowledgments}
We thank an anonymous referee for his detailed report which has helped improve the
paper significantly, and Neeraj Gupta, Nirupam Roy 
and Dave Strickland for their comments
on the manuscript. AH thanks the Kanwal Rekhi Career Development Scholarship
for partial financial support.
The GMRT is a national facility operated by the National Centre for Radio 
Astrophysics of the Tata Institute of Fundamental Research. VLA is a operated 
by Associated Universities, Inc. under contract with the National Science 
Foundation. This research has made use of the NASA/IPAC extragalactic database 
(NED) which is operated by the Jet Propulsion Laboratory, Caltech, under 
contract with the National Aeronautics and Space Administration. \\


\begin{thebibliography}{}
\bibitem[]{} Armus L., Heckman T.M., Miley G.K.,  1988, ApJ, 326L, 45
\bibitem[]{} Baars J.W.M., Genzel R., Pauliny-Toth I.I.K, Witzel A., 1977, A\&A, 61, 99
\bibitem[]{} Baum S.A., O`Dea C.P., Dallacassa D., de Bruyn A.G., Pedlar A., 1993. ApJ, 419, 553
\bibitem[]{} Bridle A. H., Perley R. A., 1984, ARA\&A, 22, 319
\bibitem[]{} Carilli C.L., Holdaway M.A., Ho P.T.P., de Pree C.G., 1992, ApJ, 399L, 59
\bibitem[]{} Cecil G., 1988, ApJ, 329, 38
\bibitem[]{} Cecil G., Bland J., Tully R.B., 1990, ApJ, 355, 70
\bibitem[]{} Cecil G., Bland-Hawthorn J., Veilleux S., Filippenko A.V., 2001, ApJ, 555, 338
\bibitem[]{} Chapman S.C., Morris S. L., Alonso-Herrero A., Falcke H., 2000, MNRAS, 314, 263
\bibitem[]{} Christopoulou P. E., Holloway A. J., Steffen W., Mundell C. G., Thean A. H. C., Goudis C. D., 
             Meaburn J., Pedlar A.,  1997, MNRAS, 284, 385
\bibitem[]{} Cid Fernandes R., Heckman T., Schmitt H., Delgado R. M. G., Storchi-Bergmann T.,  2001, ApJ, 558, 81
\bibitem[]{} Clements E.D., 1981, MNRAS, 197, 829
\bibitem[]{} Colbert E.J.M., Baum, S.A., Gallimore J.F., O'Dea C.P., Christensen J.A., 1996, ApJ, 467, 551 
\bibitem[]{} Collison P.M., Saikia D.J., Pedlar A., Axon D.J.,  Unger S.W., 1994, MNRAS, 268, 203
\bibitem[]{} Condon J.J., 1992, ARA\&A, 30, 575
\bibitem[]{} Condon J.J., Condon M. A., Gisler G., Puschell J. J., 1982, ApJ, 252, 102
\bibitem[]{} Condon J.J., Helou G., Sanders D.B., Soifer B.T., 1990, ApJS, 73, 359
\bibitem[]{} Condon J.J., Helou G., Sanders D. B., Soifer B. T., 1996, ApJS, 103, 81
\bibitem[]{} Condon J.J., Yin Q.F., Thuan T.X., Boller Th., 1998, AJ, 116, 2682
\bibitem[]{} Conti P.S.,  1991, ApJ, 377, 115
\bibitem[]{} Dahlem M., Aalto S., Klein U., Booth R., Mebold U., Wielebinski R., Lesch H., 1990, A\&A, 240, 237
\bibitem[]{} Eckart  A., Cameron M., Jackson J. M., Genzel R., Harris A. I., Wild W., Zinnecker H.,  1991, ApJ, 372, 67
\bibitem[]{} Eckart A., Cameron M., Boller Th., Krabbe A., Blietz M., Nakai N., Wagner S. J., Sternberg A.,  1996, ApJ, 472, 588
\bibitem[]{} Elmouttie M., Koribalski B., Gordon S., Taylor K., Houghton S., Lavezzi T., Haynes R., 
             Jones K., 1998a, MNRAS, 297, 49
\bibitem[]{} Elmouttie M., Haynes R. F., Jones K. L., Sadler E. M., Ehle M., 1998b, MNRAS, 297, 1202
\bibitem[]{} Forbes D.A., Boisson C., Ward M.J., 1992, MNRAS, 259, 293
\bibitem[]{} Ford H. C., Crane P. C., Jacoby G. H., Lawrie D. G., van der Hulst J. M.,  1985, ApJ, 293, 132
\bibitem[]{} Gallimore J.F., Axon D.J.,  O'Dea C.P., Baum S.A., Pedlar A., 2006, AJ, 
             in press (astro-ph/0604219)
\bibitem[]{} Garc\'{i}a-Barreto J.A., Franco J., Rudnick L., 2002, AJ, 123, 1913
\bibitem[]{} Garc\'{i}a-Barreto J. A., Scoville N. Z., Koda J., Sheth K., 2005, AJ, 129, 125
\bibitem[]{} Gull S.F., Northover K.J.E., 1973, Nature, 244, 80
\bibitem[]{} Heckman T.M., Armus L., Miley G.K., 1990, ApJS, 74, 833
\bibitem[]{} Heckman T.M., Lehnert M.D., Strickland D.K., Armus L., 2000, ApJS, 129, 493
\bibitem[]{} Hota A., Saikia D.J., 2005, MNRAS, 356, 998
\bibitem[]{} Irwin J.A., Saikia D.J., 2003, MNRAS, 346, 977
\bibitem[]{} Irwin J.A., Seaquist E.R., 1988, ApJ, 335, 658
\bibitem[]{} Jim\'{e}nez-Bail\'{o}n E., Santos-Lle\'{o} M., Dahlem M., Ehle M., Mas-Hesse J.M., Guainazzi M., 
             Heckman T.M., Weaver K.A., 2005, A\&A, 442, 861
\bibitem[]{} Jogee S., Kenney J. D. P., Smith B.J., 1998, ApJ, 494, L185
\bibitem[]{} Kewley L.J., Heisler C.A., Dopita M.A., Sutherland R., Norris R.P., Reynolds J., 
             Lumsden S., 2000, ApJ, 530, 704
\bibitem[]{} Kinney A. L., Schmitt H. R., Clarke C. J., Pringle J. E., Ulvestad J. S., Antonucci R. R. J.,  2000, ApJ, 537, 152
\bibitem[]{} Kohno K., Matsushita S., Vila-Vilar\'{o} B., Okumura S. K., Shibatsuka T., Okiura M., 
             Ishizuki S., Kawabe R., 2001, in The Central Kiloparsec of Starbursts and AGN: 
             The La Palma Connection, eds Knapen J.H., Beckman J.E., Shlosman I., Mahoney T.J., 
             ASP Conf. Proc., 249, 672
\bibitem[]{} Kondratko P.T., Greenhill L.J., Moran  J.M., 2005, ApJ, 618, 618 
\bibitem[]{} Leahy J.P., Muxlow T.W.B., Stephens P.W., 1989, MNRAS, 239, 401 	
\bibitem[]{} Leon S., Eckart A., Laine S., Schinnerer E., 2003, in Active Galactic Nuclei: 
             from Central Engine to Host Galaxy, eds S. Collin, F. Combes and I. Shlosman., 
             ASP Conf. Proc., 290, 395
\bibitem[]{} McDonald A.R., Muxlow T.W.B., Wills K.A., Pedlar A., Beswick R.J., 2002, MNRAS, 334, 912 	
\bibitem[]{} Mohan N.R., Anantharamaiah K.R., Goss W.M., 2002, ApJ, 574, 701
\bibitem[]{} Muxlow T.W.B., Pedlar A., Wilkinson P.N., Axon D.J.,  Sanders E.M., de Bruyn A.G.,  1994, 
             MNRAS, 266, 455
\bibitem[]{} Osterbrock D. E., Cohen R. D., 1982, ApJ, 261, 64
\bibitem[]{} Pogge R.W., 1988, ApJ, 328, 519
\bibitem[]{} Reuter H.-P., Klein U., Lesch H., Wielebinski R., Kronberg P.P., 1992, A\&A, 256, 10 	
\bibitem[]{} Rubin V. C., Thonnard N., Ford W. K., Jr., 1975, ApJ, 199, 31
\bibitem[]{} Saikia D.J., Unger S.W., Pedlar A., Yates G.J., Axon D.J., Wolstencroft R.D., Taylor K., 
             Gyldenkerne K., 1990, MNRAS, 245, 397
\bibitem[]{} Sakamoto K., et al., 2006, ApJ, 636, 685
\bibitem[]{} Schinnerer E., Eckart A., Boller Th.,  2000, ApJ, 545, 205
\bibitem[]{} Seaquist E. R., Odegard N., 1991, ApJ, 369, 320
\bibitem[]{} Spergel D.N. et al., 2003, ApJS, 148, 175
\bibitem[]{} Stone J.L. Jr., Wilson A.S., Ward M.J., 1988, ApJ, 330, 105
\bibitem[]{} Strickland D.K., Heckman T.M., Colbert E.J.M., Hoopes C.G., Weaver K.A., 2004, 
             ApJS, 151, 193 
\bibitem[]{} Ulvestad J.S., Antonucci R.R.J., 1997, ApJ, 488, 621
\bibitem[]{} Ulvestad J.S., Wilson A.S., 1989, ApJ, 343, 659
\bibitem[]{} Ulvestad J.S., Wilson A.S., 1984, ApJ, 285, 439
\bibitem[]{} Ulvestad J.S., Wilson A.S., Sramek R.A., 1981, ApJ, 247, 419
\bibitem[]{} Veilleux S., 2001, in Starburst Galaxies: Near and Far, 
             eds Tacconi L., Lutz D., Springer-Verlag, Heidelberg, p. 88
\bibitem[]{} Veilleux S., Rupke D.S., 2002, ApJ, 565, L63
\bibitem[]{} Veilleux S., Shopbell P. L., Miller S. T., 2001, AJ, 121, 198
\bibitem[]{} Veilleux S., Cecil G., Bland-Hawthorn J.,  2005, ARA\&A, 43, 769
\bibitem[]{} Webber W.R., 1991, in The interpretation of modern synthesis observations of 
             spiral galaxies, eds Duric N., Crane P.C., ASP Conf. Series, 18, 37
\bibitem[]{} Wilcots E.M., Turnbull M.C., Brinks E., 2001, ApJ, 560, 110
\bibitem[]{} Wills K.A., Pedlar A., Muxlow T.W.B., Stevens I.R., 1999, MNRAS, 305, 680	
\bibitem[]{} Wilson A.S., Willis A. G, 1980, ApJ, 240, 429
\bibitem[]{} Wilson A.S., Wu Xuening, Heckman T. M., Baldwin J. A., Balick B., 1989, ApJ, 339, 729
\bibitem[]{} Yoshida M. Taniguchi Y., Murayama T., 1999, AJ, 117, 1158
\bibitem[]{} Zurita A., Rozas M., Beckman J. E.,  2000, A\&A, 363, 9

\end{thebibliography}
\end{document}